\documentstyle[prl,aps,multicol,epsf]{revtex}
\begin{document}
\tighten
\draft

\title{Flux-Line Lattice Structures in Untwinned YBa$_2$Cu$_3$O$_{7-\delta}$}

\author{S. T. Johnson$^{1,7}$, E. M. Forgan$^1$, S. H. Lloyd$^1$, 
C. M. Aegerter$^2$, S. L. Lee$^3$, R. Cubitt$^4$, P. G. Kealey$^1$, C. Ager$^3$,
S. Tajima$^5$, A. Rykov$^5$, and D. McK. Paul$^6$}

\address{$^1$School of Physics and Astronomy, University of Birmingham,
Birmingham B15 2TT, United Kingdom.}
\address{$^2$Physik-Institut der Universit\"{a}t Z\"{u}rich, CH-8057 
Z\"{u}rich, Switzerland.}
\address{$^3$School of Physics and Astronomy, University of St. Andrews,
St. Andrews KY16 9SS, Scotland, United Kingdom.}
\address{$^4$Institut Max von Laue-Paul Langevin, 38042 Grenoble, France.}
\address{$^5$Superconductivity Research Laboratory, ISTEC, Tokyo 135, Japan}
\address{$^6$Department of Physics, University of Warwick, Coventry
CV4 7AL.  United Kingdom.}
\address{$^7$Laboratoire de Physique des
Solides, B\^{a}timent 510, Universit\'{e} Paris-Sud, 91405 Orsay,
France.}
\date{January 12th, 1999}

\maketitle

\begin{abstract}
A small angle neutron scattering study
of the flux-line lattice in a large single
crystal of untwinned YBa$_2$Cu$_3$O$_{7-\delta}$ is presented. 
In fields parallel to the {\it c}-axis, diffraction spots
are observed corresponding to four orientations of a hexagonal lattice,
distorted by the {\it a-b} anisotropy. A value for the anisotropy,
the penetration depth ratio, of $\lambda_a/\lambda_b$=1.18(2) 
was obtained. The high quality of the data is such that 
second order diffraction is observed, indicating a well
ordered FLL. With the field at 33$^\circ$ to {\it c} a field dependent 
re-orientation of the lattice is observed around 3T.
\end{abstract}


\begin{multicols}{2}
\narrowtext

The remarkable properties of the mixed state in the cuprate high-T$_c$
superconductors are of great current interest.
In particular, the expectation that
high-T$_c$ superconductors have an unconventional 
pairing symmetry has led inevitably to the question: how does 
the structure of the flux-line lattice (FLL)
differ between conventional and unconventional
superconductors? The question has been taken up by several recent
theoretical contributions \cite{theory,dwave2,franz,chang} which
predict a variety of interesting FLL effects all deviating 
from the benchmark triangular Abrikosov lattice.  
However, such discussions may presuppose that the crystallographic
properties of the FLL are already well understood in the simplest 
low-field regime where unconventional effects are least prevalent.
This has been far from the experimental truth. Observations require a probe
sensitive to the microscopic arrangement of flux-lines. 
Direct imaging \cite{maggio} and 
decoration techniques \cite{dolan} all have inherent drawbacks,
and muon spin rotation has not yet achieved 
the sophistication to resolve the most subtle effects \cite{sonier}. 
By comparison, small angle neutron scattering (SANS)  
provides unrivalled insights 
into the crystallography of the FLL, and is
the only technique capable of unequivocably resolving such
questions.

The demanding nature of neutron experiments requires large
single crystals (masses $\gtrsim$ 200mg), and because of
this YBa$_2$Cu$_3$O$_{7-\delta}$ (YBCO) has been the 
cuprate of choice for SANS experiments \cite{bscco}.
However, the materials properties
of YBCO are complicated. The presence of Cu-O 
chains which are aligned with the crystallographic 
{\it b} direction render the otherwise tetragonal structure
orthorhombic. Upon cooling from the growth, 
twin boundaries form along \{110\} directions
separating domains of interchanged {\it a} and {\it b} axes. 
A strong interaction between flux-lines and these twin planes
significantly influences FLL properties. 
There is a further effect of the chains. Although the orthorhombic
distortion is only slight ($\approx$ 1\%), the electronic
structure is markedly affected, and
the consequence is anisotropy within the 
{\it ab}-plane of both superconducting \cite{pollight,sun}
and normal state \cite{friedmann} properties. All
previous SANS studies have been on heavily twinned crystals
\cite{keimer,mona3019}, and 
although observations of a pattern with four-fold symmetry
were claimed to be due to unconventional d$_{x^2-y^2}$ pairing \cite{keimer}, 
it could not be discounted that alignment by twin planes,
in combination with the {\it a-b} anisotropy, did not provide a 
more plausible explanation \cite{reply}. 

A new development in materials technology, involving the 
application of uniaxial stress during the cooling 
process, has made possible the 
growth of very large untwinned crystals \cite{rykov}.
In this paper we report SANS measurements 
on such untwinned single crystals of YBCO. The results
show the effects of the {\it a-b} anisotropy on the FLL,
and prove explicitly that the results from twinned crystals 
\cite{keimer} cannot be interpreted as evidence
for d-wave effects. Having clarified the low-field picture,
we then present the first indication at higher
fields that the range of validity
of simple London scaling theory is limited.

The structure of the lattice formed by flux-lines in a superconductor
is dictated in crude terms, as a function
of flux-line density, by the intrinsic anisotropy of the 
electronic structure and 
the shape of the core of individual flux-lines, as well as by
the extrinsic effects of flux-line pinning by defects and surfaces. 
The penetration depth, $\lambda$, and coherence length, $\xi$, 
being the parameters which define
the two length scales of flux-line interactions, their ratio
$\kappa=\lambda/\xi$ is naturally an important quantity. Where $\kappa$ 
is not much bigger than unity, core interactions exist over the
whole (H,T) space and structures ranging from triangular to square
are observed, as in borocarbides \cite{donbc} and Nb \cite{lowk},
as a manifestation of anisotropic electronic structures 
dependent upon the orientation of the applied field with respect
to the atomic lattice. In high-$\kappa$ superconductors, 
such as cuprates and low-T$_c$ NbSe$_2$ \cite{nbse}, 
one would expect that core effects only become
influential at fields comparable with H$_{c2}$, and at low
enough fields a purely electrodynamic interaction 
model should be sufficient. 
In a basic London model, incorporating anisotropy in
the effective mass tensor, predictions for fields applied
parallel to the principal 
{\it c}-axis are a hexagonal
FLL which is degenerate in energy with respect to 
orientation with the atomic lattice 
\cite{campbell:2439}. In reality, the presence of
even vanishingly small higher-order effects, 
intrinsic or extrinsic in origin, will be
able to lift this degeneracy to produce a preferred 
orientation. Among intrinsic effects, it is in 
the structure of the vortex core that an unconventional pairing symmetry is
revealed, calculations predicting it to be four-fold 
\cite{dwave2}. Free energy calculations have therefore been made 
incorporating four-fold symmetry through higher gradient
terms in a GL-type theory \cite{theory,chang}, and in  
a complete microscopic derivation of non-local
electrodynamics in a London model \cite{franz}. 
The predictions are for a distorted hexagonal
lattice at low field transforming to a square lattice over a
field range which is dependent upon adjustable parameters 
related to the strength of the d$_{x^2-y^2}$ contribution. 

The measurements have been obtained 
over a series of experiments using
three SANS instruments (D11, D17 and D22)
at the Institut Laue-Langevin, Grenoble, France. 
A field was applied parallel to the
incident neutron beam, and the FLL formed by cooling through T$_c$.
Rotation of the sample about a vertical axis with respect to the
field, and of the field and sample as one with respect to the
beam were both possible. As in all static imaging techniques, 
we see the FLL structures
"frozen in" at the irreversibility line near T$_c$.
Two crystals were initially studied, of masses 312mg and 1125mg,
grown and detwinned under the same conditions, and oxygenated 
in flowing O$_2$ at 490$^\circ$C for optimal doping. 
Magnetization measurements showed T$_c$ of 92K.
Essentially identical results were obtained from both, 
and the work presented therefore relates solely 
to the larger.

\begin{figure}
\centerline{\epsfxsize=9.0cm\epsfbox{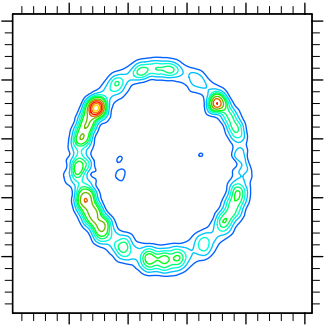}}
\caption{
The SANS pattern with 0.51T applied
parallel to the {\it c}-axis; {\it a}-axis is vertical in the figure.
The plot area covers 128x128 detector pixels, 
in reciprocal space $\pm$0.0173$\AA^{-1}$; neutron wavelength 
$\lambda_n$=12$\AA$, detector distance 14.5m, and collimation 14.4m.
}
\label{fig0}
\end{figure}

In Figure \ref{fig0}, a diffraction pattern is shown, obtained at 1.5K
with a relatively low field of 0.51T applied parallel to the {\it c}-axis. 
The diffraction spots lie in an elliptical ring, representing a FLL
which is essentially polycrystalline. This is in distinct 
contrast to patterns obtained for
twinned crystals \cite{keimer,mona3019} which show four-fold patterns.
The ellipse shape is clearly due to the superconducting
anisotropy between the {\it a} and {\it b} basal plane directions.
For this to be observable requires that the 
bulk of the sample must have a single
Cu-O chain orientation. 
For fields larger than B$_{c1}$ \cite{daemen}, the axial ratio of the
ellipse is equal to the ratio of the penetration depths,
$\gamma_{ab}$ =  $\lambda_a/\lambda_b$ 
(making $\gamma^2 = m_a/m_b$ the effective mass ratio).
The sign of the anisotropy
is consistent with a reduced penetration depth $\lambda_b$
due to the increased supercurrent flow along the chain direction.
From fits to the ellipse of scattering, 
a value for $\gamma$ of 1.18(2) was obtained. This is
slightly higher than the 1.13 estimated by 
Bitter decoration \cite{dolan}, but in the lower range
of values found by polarized reflectivity, $\gamma$=1.3-1.6
\cite{pollight}, and Josephson tunnel junctions, $\gamma$=1.2-1.7
\cite{sun}. That the variability is between samples
rather than techniques is borne out by $\mu$SR
measurements on the same sample which give $\gamma$=1.16(2) \cite{cameron}.
No field dependence in $\gamma$ was observed. 
Reflectivity measurements on strongly oxygen disordered 
samples give  $\gamma$ as low as 1.05
\cite{schlesinger}. We conclude that the Cu-O chains
in our crystal are slightly disrupted by a small concentration of 
oxygen vacancies or impurities.

\begin{figure}
\centerline{\epsfxsize=6cm\epsfbox{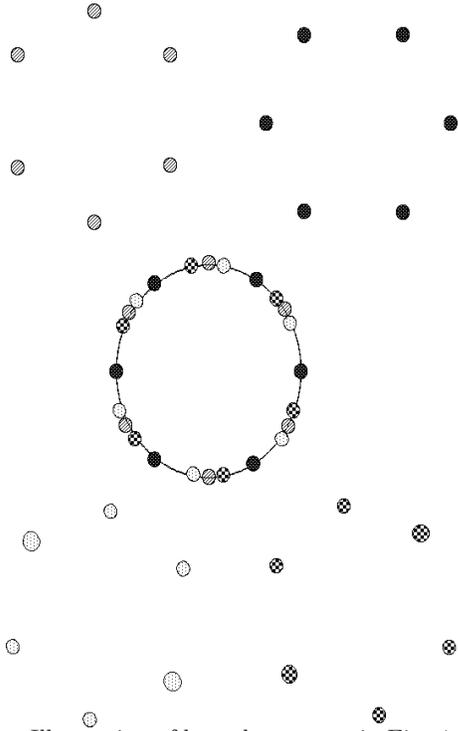}}
\caption{
Illustration of how the pattern in Fig. \ref{fig0}
is formed from two reciprocal lattices having planes 
oriented with the primary atomic lattices (top), and two having planes aligned 
with the \{110\} and \{1$\bar{1}$0\} twin planes (bottom) producing spots
at 45$^\circ$ (emphasised).
}
\label{spots}
\end{figure}

The intensity distribution around the ring in Fig. \ref{fig0}, 
although unbroken, contains distinct diffraction spots. 
From the assessment of many such patterns, we conclude
that four different flux-line lattice orientations are being
observed, each contributing a hexagonal pattern of six spots
distorted by the {\it a-b} anisotropy. The situation is illustrated
in Fig \ref{spots}. Two of the structures have a FLL plane oriented
with the axes of the atomic lattice, and two have a plane at
45$^\circ$ to them (that is with the \{110\} directions).
Surprisingly, the latter two are the same as two of the four
orientations observed in twinned crystals. 
The implication is that this crystal is not completely free of twins,
although the domains of the opposite orientation must be
a very small fraction of the volume otherwise a second
set of spots aligned on an ellipse at right angles would
be observed. A neutron study of the crystallography
of these YBCO crystals has confirmed this
\cite{kealey}, the minority orientation was found to be confined 
to less than $5 \%$ of the sample.
The two 45$^\circ$ distorted hexagonal patterns correspond to part of
the square pattern observed by Keimer {\it et al} \cite{keimer}, 
who claimed that the orientation and distortion were
due to d-wave effects. It is made clear by the present observations 
that the FLL structures, instead, are controlled at these low fields by
anisotropy plus twin plane pinning in the manner postulated
earlier \cite{reply}. Although a non-local contribution to the
penetration depth anisotropy is possible \cite{kogan}, the
{\it field-dependence} of the flux latice distortion argues
against this being important.

\begin{figure}
\centerline{\epsfxsize=10.0cm\epsfbox{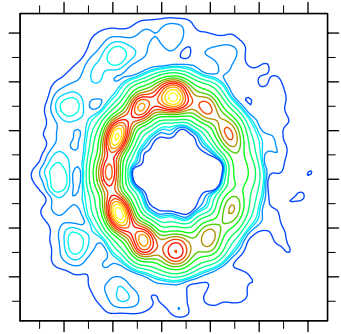}}
\caption{
The pattern with a field
of 0.2T, not quite parallel to c. A ring of 
twelve first-order spots from two hexagonal lattices is visible, 
and remarkably well defined second-order spots. The plot covers
64x64 detector pixels, $\pm$0.0144$\AA^{-1}$ in reciprocal space;
$\lambda_n$=10$\AA$, detector distance 14m, and collimation 13.6m. 
}
\label{fig1}
\end{figure}

By tilting the {\it c}-axis $\approx$ 1.5$^\circ$ (about the vertical) 
away from the applied field direction the different nature of the FLLs
is made apparent. As expected of flux-lines pinned
along their length by extended defects, the two 45$^\circ$ FLLs remain
fixed to the twin plane direction while the remaining two lie instead
along the field direction. Consequently in Fig. \ref{fig1}, this remarkably 
clear pattern shows only the twelve first order spots of the lattices
oriented with the {\it a} and {\it b} axes. A ring of second order 
reflections is testament  that the FLL is well formed.
The left-hand side is stronger because these spots have been brought 
closer to their Bragg condition by rotating the
field and sample as one. The Bragg condition for spots of the 45$^\circ$ 
FLLs are separated a further 1.5$^\circ$ away, which is considerably larger 
than the rocking curve widths. Gaussians fits to individual spots
gave full-widths at half-maximum of 0.65(5)$^\circ$ and 0.71(5)$^\circ$ 
for the {\it a-b} and 45$^\circ$ FLLs respectively (instrumental
broadening contributes only 0.2$^\circ$).
These give a lower limit for the longitudinal correlation lengths
of the flux-lines of 1.8${\mu}$m.

The intensity of a reflection integrated over its full
rocking curve can be related to the penetration
depth. The established expression for the 
structure factor derived
from London theory \cite{christen:4506} indicates that the
intensity is proportional to $\lambda^{-4}$. Making this
calculation for the rocking curve of the pattern in Fig. \ref{fig1}
requires that we estimate the proportion of the total number
of flux-lines contained in each lattice.
The rocking curves give (within 10$\%$) equal proportions of 
the intensity distributed between them. Using this, we obtain
$\lambda$=138(5)nm. If we take  
$\lambda$= $\sqrt{\lambda_a \lambda_b}$, and use
the already measured value of $\gamma$, 
$\lambda_a$=150(6)nm and  $\lambda_b$=127(6)nm are obtained.
The value of $\lambda_a$ provides a useful check of the 
consistency of our results, as it is known to be insensitive
to the state of the chain layers, and is therefore
reproducible across samples. It is typically quoted as
155nm for optimal doping \cite{tallon} which agrees 
within experimental error.

Over the range of fields investigated, up to a maximum
of 4 Tesla, the FLL struture remained essentially unchanged.
This may not be inconsistent with the various unconventional and 
non-local theories whose distortions, if present, may simply be
too weak over this range.
We note that the distorted square lattices observed by STM at higher
fields \cite{maggio} are not consistent with our observations.
Our first indication of the breakdown of simple London scaling
theory comes from measurements made with the field inclined
at an angle to the {\it c}-axis.

\begin{figure}
\centerline{\epsfxsize=9.2cm\epsfbox{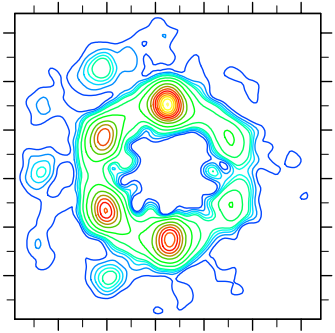}}
\centerline{\epsfxsize=9.3cm\epsfbox{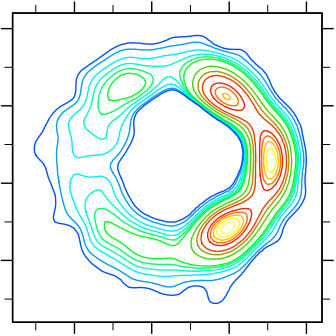}}
\caption{
The applied field tilted towards the {\it b}-axis at an angle of 
33$^\circ$ to the {\it c}-axis.
At 0.2T in the upper pattern a single 
hexagonal lattice aligned with the {\it a}-axis 
(vertical in the plots) is observed;
$\pm$0.0144$\AA^{-1}$, $\lambda_n$=10$\AA$, 
detector distance 14m, and collimation 13.6m. At fields above 3T, the lower 
pattern changes to be oriented instead with the {\it b}-axis;
$\pm$0.0503$\AA^{-1}$,
$\lambda_n$=10$\AA$, detector distance and collimation are 2.5m.
}

\label{fig3}
\end{figure}

Rotating the field away from {\it c} to large angles, 
another factor becomes important:
the anisotropy between supercurrents along {\it c}
compared with those in the {\it ab}-plane. 
Previous SANS measurements on twinned crystals have given 
$\gamma_c =\lambda_c /\lambda_{ab}\approx$4.5 \cite {mona3019}.
This $\gamma_c$ anisotropy mixes with the $\gamma_{ab}$ anisotropy. 
With {\it b} as the vertical axis of rotation
the eccentricity of the ellipse is accentuated by $\gamma_c$  adding 
to $\gamma_{ab}$,
and at angles comparable to 30${^\circ}$ from {\it c} a continuous 
highly eccentric ellipse of
scattering is observed, indicating a polycrystalline FLL.
In contrast, when {\it a} is the axis of rotation, spots from the 
two {\it ab} aligned FLLs remain, with the
{\it a}-oriented FLL becoming dominant with increasing angle.
This is the orientation expected from simple London theory 
\cite{campbell:2439}.
A possible explanation for the difference rotating about the two
axes, is that
rotation about {\it b} leaves the chains perpendicular to the applied
field, while rotation about {\it a} leads to a shallower angle
between field and chain direction, suggesting an
unexpected influence of the chain structure upon the flux-lines.
At an angle of $\approx$33${^\circ}$ from {\it c},
shown in Fig. \ref{fig3}(a), the 
two anisotropies cancel, and an
almost undistorted hexagonal pattern is observed, 
again with second order spots.
The FLL is {\it a}-axis oriented; however, under the same conditions
but at a field of 3.8T, the pattern in Fig. \ref{fig3}(b) is 
observed corresponding to a FLL oriented instead
with the {\it b}-axis. There is clearly a field dependence to whatever
mechanism is controlling the FLL orientation, or perhaps a
crossover from the London mechanism at low flux-line density to another
at high density. 
At present we have no explanation for the effect, and it may yet
prove necessary to invoke core effects, perhaps of $d_{x^2-y^2}$ origin,
to account for it. 

In conclusion, we emphasise that the crystallography of the FLL
must be considered in the context of a complicated variety
of influences, before addressing the
singular question of unconventional pairing.
The results demonstrate the various implications for the FLL
caused by the presence of CuO chains in YBCO; introducing
microstructural complications of twinning, a basal plane anisotropy
in the effective mass, and even the suggestion of an interaction
between the flux-lines and the chain structure. To resolve
whether the observed field dependant reorienation is an intrinsic 
feature of unconventional superconductivity will require experiments 
on cuprates free of chains and ideally with tetragonal symmetry.

Financial support was provided by the Engineering and
Physical Sciences Research Council.
The crystal growth work was supported by NEDO Japan for the
R\&D of Industrial Science and Technology Frontier Program.
STJ is funded by a Marie Curie Fellowship of the EC.

\end{multicols}
\end{document}